\documentclass[12pt]{article}
\usepackage{graphicx,amsmath}

\newlength{\dinwidth}
\newlength{\dinmargin}
\def\lapproxeq{\lower .7ex\hbox{$\;\stackrel{\textstyle                                                    
<}{\sim}\;$}}                                                    
\def\gapproxeq{\lower .7ex\hbox{$\;\stackrel{\textstyle                                                    
>}{\sim}\;$}}                                                    
\def\be{\begin{equation}}                                                    
\def\ee{\end{equation}}                                                    
\def\bea{\begin{eqnarray}}                                                    
\def\eea{\end{eqnarray}}

\begin{document}                                                    
\titlepage                                                    
\begin{flushright}                                                    
IPPP/09/43  \\
DCPT/09/86 \\                                                    
\today \\                                                    
\end{flushright}                                                    
                                                    
\vspace*{0.5cm}                                                    
                                                    
\begin{center}                                                    
{\Large \bf On minimum-bias effects at the LHC}                                                                                                        
                                                    
\vspace*{1cm}                                                    
V.A. Khoze$^{a,b}$, A.D. Martin$^a$ and M.G. Ryskin$^{a,b}$  \\                                                    
                                                   
\vspace*{0.5cm}                                                    
$^a$ Institute for Particle Physics Phenomenology, University of Durham, Durham, DH1 3LE \\                                                   
$^b$ Petersburg Nuclear Physics Institute, Gatchina, St.~Petersburg, 188300, Russia            
\end{center}                                                    
                                                    
\vspace*{1cm}                                                    
                                                    
\begin{abstract}

In general, minimum-bias triggers planned for experiments at the LHC miss a considerable fraction of the total number of events. We exemplify the rejection rate using the Durham model of soft high-energy interactions to obtain quantitative estimates of the signals arising from the ATLAS scintillation counters positioned in the rapidity intervals $2<|\eta|<4$, and also from TOTEM detectors covering the intervals $3.1<|\eta|<6.5$.  Typically we find that the expected signal is about half of the total cross section, $\sigma_{\rm tot}$. We also calculate the cross section for the so-called zero-bias measurement planned by CMS in the extended rapidity interval $-5<\eta<7$. We emphasize that only models which give satisfactory predictions for the measured minimum-bias or zero-bias cross sections can be used to obtain the value of the total inelastic cross section.

\end{abstract}    

In hadron collider experiments we do not have the possibility to study all the event topologies produced in the $pp$ (or $p\bar{p}$) interactions. In general, experiments collect only a fraction of the events, selected by one or another trigger. Even the sample of so-called minimum-bias events is restricted by some minimum-bias trigger condition. 

It is informative to discuss the situation in terms of realistic examples. For illustration, we study the minimum-bias effect in the ATLAS experiment at the LHC, where inelastic events produce at least one charged particle in scintillation counters\footnote{These counters are known as minimum-bias trigger scintillators (MBTS) \cite{MBTS}.} which cover the rapidity intervals $2<|\eta|<4$. We also consider the analogous minimum-bias trigger that is possible using the T1 and T2 detectors of TOTEM \cite{TOTEM}, which cover the rapidity intervals $3.1<|\eta|<6.5$.

 At first sight, it appears that almost all the inelastic events should satisfy these trigger conditions. However, detailed theoretical computations show that these `minimum-bias' triggers reject up to about 25-30\% of the total inelastic cross section, $\sigma_{\rm inel}=\sigma_{\rm tot}-\sigma_{\rm el}$. Here we present estimates of the different contributions which miss such a trigger. Of course, such estimates are model dependent. Therefore, it is better not to extrapolate the measured signal, using one or another model, to obtain $\sigma_{\rm tot}$, but instead to compare the data directly with the predictions for the cross section selected by the trigger. 

Let us list the processes which do {\it not} produce particles in the rapidity intervals covered by the trigger. First, we have elastic scattering events. Next, we have events caused by low-mass single or double proton dissociation. Events can also be due to high-mass dissociation with gaps at the trigger rapidity intervals. Then, there may be events with two or more rapidity gaps like central diffractive production; that is, `so-called' double-Pomeron exchange (DPE) production, see, for example, \cite{soft}. However, in realistic models this last contribution is rather small. Finally, there may be just a fluctuation in a normal inelastic event which produces no secondaries in the trigger rapidity interval.

Both the ATLAS scintillation counters and the TOTEM detectors cover `symmetric' rapidity intervals, $\eta_1<|\eta|<\eta_2$. For each experiment  we can consider three different minimum-bias triggers:
\bea
{\rm MBT1}: & &{\rm only}~\eta_1<\eta<\eta_2~{\rm trigger~fired} \label{eq:MBT1}\\
{\rm MBT2}: & &{\rm either}~\eta_1<\eta<\eta_2~{\rm or}~-\eta_2<\eta<-\eta_1~{\rm triggers~fired} \label{eq:MBT2} \\
{\rm MBT3}: & &{\rm trigger~requires~events~in~}both~ (\eta_1,\eta_2) ~{\rm and}~ (-\eta_2,-\eta_1)~ {\rm intervals} \label{eq:MBT3}.
\eea

\begin{figure} [t]
\begin{center}
\includegraphics[height=4cm]{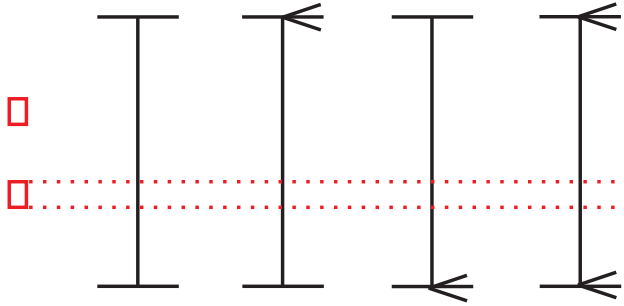}
\caption{\sf Schematic diagrams of the quasi-elastic events which do not fire either of the $\eta_1<|\eta|<\eta_2$ minimum-bias counters. The counters are shown as small rectangles along the rapidity axis.}
\label{fig:low}
\end{center}
\end{figure}
To predict the size of the contributions of the various processes listed above, for the three triggers, it is important to choose a model of soft high-energy interactions which describes both low- and high-mass dissociation of the proton, as well as elastic scattering.
We take the multichannel eikonal, multi-Pomeron model given in Ref. \cite{soft,KMR}, which was tuned to describe all the available `soft' data in the CERN-ISR to Tevatron energy range. The model predicts that the total and elastic scattering cross section are $\sigma_{\rm tot}=91.5$ mb and $\sigma_{\rm el}= 21.5$ mb at $\sqrt{s}=14$ TeV. If we include the possibility of low-mass excitations of one or both protons together with the elastic cross section, then the model gives $\sigma_{{\rm low}M}=26.7$ mb. This is the cross section of quasi-elastic events which do {\it not} produce a signal for any of the minimum-bias triggers, see Fig.~\ref{fig:low}. 

\begin{figure} 
\begin{center}
\includegraphics[height=7cm]{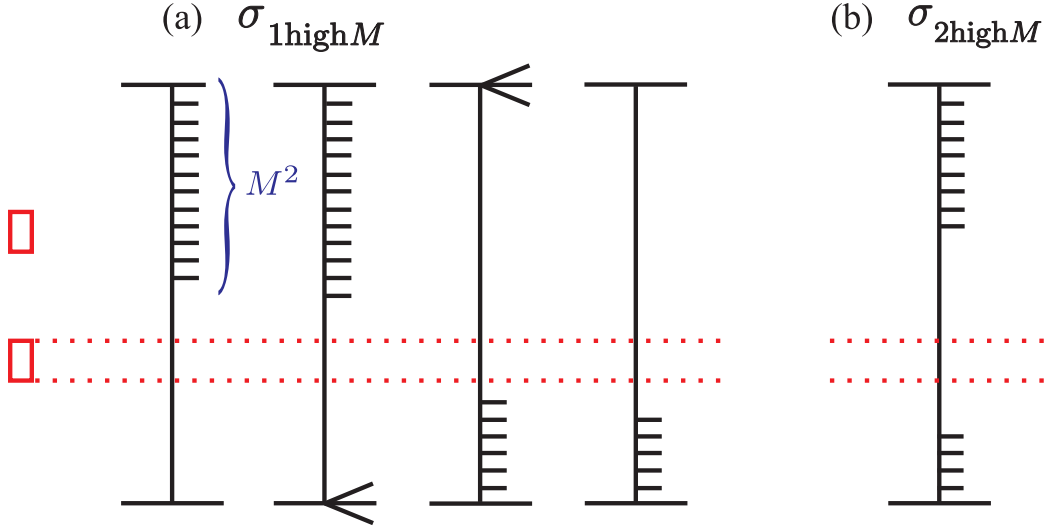}
\caption{\sf Schematic diagrams of events containing (a) one, and (b) both, protons dissociating into a high-mass system, which do not trigger the scintillation counter in the $2<\eta<4$ rapidity interval.  Both counters are shown as small rectangles. }
\label{fig:high}
\end{center}
\end{figure}

We discuss the other processes that may be missed by the minimum-bias triggers, (\ref{eq:MBT1})$-$(\ref{eq:MBT3}), in terms of the ATLAS scintillation counters, which cover the $2<|\eta|<4$ rapidity intervals. For the moment, we take the LHC energy to be $\sqrt{s}$=14 TeV. In addition to the quasi-elastic events, there may be high-mass excitations of {\it one} of the protons which do not fire one or other or both of the minimum-bias triggers. We denote the corresponding cross sections as $\sigma_{1{\rm high}M}$, see Fig.~\ref{fig:high}(a).  Detailed model calculations give cross sections of 16.7 mb, 13.1 mb and 20.4 mb,  corresponding to the MBT1, MBT2, MBT3 triggers respectively -- that is, to the cases when, first, there are no secondaries in the $2<\eta<4$ trigger, second, no secondaries in both intervals $(-4,-2)$ and (2,4), and, third, when there are no secondaries in either of intervals.
 Finally, there may be high-mass dissociation of {\it both} protons which do not fire one
 or other or both of the minimum-bias counters, with cross sections\footnote{$\sigma_{1{\rm high}M}$ and $\sigma_{2{\rm high}M}$ also include contributions coming from events which contain additional rapidity gaps in the interval covered by secondaries from high-mass excited states. However, these are not crucial for our computation. The important
point is that we already have a gap in the interval covered by the trigger. There may be a small correction from multi-gap events. In particular, single dissociation, which has a gap in the $(-4,-2)$ interval but, which at first sight, should fire the (2,4) counter, 
 includes DPE events, which may be missed by the MBT2 trigger
in the case when secondaries produced by Pomeron-Pomeron interactions have rapidity in the interval $|\eta|<2$. The corresponding cross section is computed to be about 0.04 mb. Thus we do not need to consider events with more rapidity gaps. This contribution is neglected from now on.} denoted by $\sigma_{2{\rm high}M}$, see Fig. \ref{fig:high}(b). In this case, the cross sections corresponding to no events for the three trigger conditions are 1.5 mb, 0.9 mb and 2.0 mb respectively. 

To calculate the cross sections {\it observed} by the three alternative minimum-bias triggers, we have to subtract from the total cross section the contributions which do not fire the respective triggers. The remainder is the signal
\be
\sigma_{\rm signal}~=~ \sigma_{\rm tot}-\sigma_{{\rm low}M}-\sigma_{1{\rm high}M}-\sigma_{2{\rm high}M}.
\label{eq:signal}
\ee
For the three triggers MBT1, MBT2 and MBT3 we have cross sections $\sigma_{\rm signal}=$ 46.6 mb, 50.8 mb and 42.4 mb respectively. That is, the minimum-bias triggers pick up only about half\footnote{Of course, it should be possible to make an accurate
measurement of the {\it elastic} cross section, for instance in the TOTEM experiment \cite{TOTEM}; so, according to the
model of Ref.~\cite{KMR}, we are only missing about 25-30\% of the total
cross section. However, here, we wish to emphasize that the
original minimum-bias measurements (and trigger conditions) should be
presented. These may be then compared directly to model
predictions. Only at this stage, if the model is in agreement
with the measurements, should extrapolations be made to
estimate the total cross section.} of the total cross section, $\sigma_{\rm tot}=91.5$ mb \cite{KMR}.

Unfortunately, the predictions for the minimum-bias triggers are at the `parton' level. To be realistic these should be corrected for {\it hadronization and detector effects}, which smear the rapidity distributions of the secondaries. These effects are not negligible, especially at the edge of the counters. In particular, a particle from a parent jet with rapidity as large as $\eta=4.5$ could fire the $2<\eta<4$ trigger.                                                 
To gain insight into the role of the `hadronization edge-effects', we repeat the computations for the extended rapidity intervals $1.5<|\eta|<4.5$. The results are shown in brackets in Table \ref{tab:T1} for collider energies $\sqrt{s}=$ 14 and 10 TeV respectively, and compared with the cross sections corresponding to the intervals $2<|\eta|<4$.

\begin{table} [h]
  \centering
  \begin{tabular}{c|ccc|ccc}
    \hline \hline
    & \multicolumn{3}{c|}{$\sqrt{s}$=14 TeV} & \multicolumn{3}{c}{$\sqrt{s}$=10 TeV} \\ \hline
     & MBT1 & MBT2 & MBT3 & MBT1 & MBT2 & MBT3 \\ \hline
$\sigma_{{\rm low}M}$ &   26.7 (26.7) & 26.7 (26.7) & 26.7 (26.7)  &   25.7 (25.7) & 25.7 (25.7) & 25.7 (25.7)    \\
$\sigma_{1{\rm high}M}$  &   16.7 (16.0) & 13.1 (12.1) & 20.4 (20.0)  &   15.7 (14.9) & 11.9 (10.8) & 19.4 (19.1)    \\ 

$\sigma_{2{\rm high}M}$  &   1.5 (1.3) & 0.9 (0.8) & 2.0 (1.8)   &   1.3 (1.2) & 0.8 (0.7) & 1.8 (1.6)    \\ \hline
$\sigma_{\rm signal}$  &   46.6 (47.5) & 50.8 (51.9) & 42.4 (43.0)   &   45.9 (46.8) & 50.2 (51.4) & 41.7 (42.2)   \\    
    \hline \hline
  \end{tabular}
\caption{\sf The cross sections (in mb), as defined in (\ref{eq:signal}) and the accompanying discussion, for the three `ATLAS' minimum-bias triggers MBT1, MBT2 and MBT3 of (\ref{eq:MBT1})$-$(\ref{eq:MBT3}) for scintillation counters with $2<|\eta|<4$. We show results for two LHC energies $\sqrt{s}$. The numbers in brackets correspond to the extended rapidity intervals $1.5<|\eta|<4.5$. At $\sqrt{s}$=10 TeV (14 TeV) the model \cite{KMR} predicts $\sigma_{\rm tot}=88.6(91.5)$ mb and $\sigma_{\rm el}=20.6(21.5)$ mb. The cross sections in the first three lines correspond to events which are outside the respective rapidity interval(s), while the last line is the cross section of the observed inelastic events.}
  \label{tab:T1}
\end{table}

The analogous predictions for the minimum-bias triggers possible with the combined T1 and T2 detectors of TOTEM, are shown in Table \ref{tab:T2}. In this case, we find that the detectors pick up a larger fraction of high-mass dissociation, particularly for the trigger MBT2.
\begin{table} [h]
  \centering
  \begin{tabular}{c|ccc|ccc}
    \hline \hline
    & \multicolumn{3}{c|}{$\sqrt{s}$=14 TeV} & \multicolumn{3}{c}{$\sqrt{s}$=10 TeV} \\ \hline
     & MBT1 & MBT2 & MBT3 & MBT1 & MBT2 & MBT3 \\ \hline
$\sigma_{{\rm low}M}$ &   26.7 (26.7) & 26.7 (26.7) & 26.7 (26.7)  &   25.7 (25.7) & 25.7 (25.7) & 25.7 (25.7)    \\
$\sigma_{1{\rm high}M}$  &   13.4 (12.1) & 5.6 (3.3) & 21.2 (20.8)  &   12.2 (10.9) & 4.1 (1.9) & 20.2 (19.9)    \\ 

$\sigma_{2{\rm high}M}$  &   0.7 (0.4) & 0.2 (0.1) & 1.2 (0.8)   &   0.5 (0.2) & 0.1 (0.03) & 0.9 (0.4)    \\ \hline
$\sigma_{\rm signal}$  &   50.7 (52.3) & 59.0 (61.4) & 42.4 (43.2)   &   50.2 (51.8) & 58.7 (61.0) & 41.8 (42.6)   \\    
    \hline \hline
  \end{tabular}
\caption{\sf The cross sections (in mb), as defined in (\ref{eq:signal}) and the accompanying discussion, for the three possible TOTEM triggers  of (\ref{eq:MBT1})$-$(\ref{eq:MBT3}), which cover the rapidity intervals $3.1<|\eta|<6.5$.  The numbers in brackets correspond to the extended rapidity intervals $2.6<|\eta|<7$. The cross sections in the first three lines correspond to events which are outside the respective rapidity interval(s), while the last line is the cross section of the observed inelastic events.}
  \label{tab:T2}
\end{table}

Next, we estimate the effect of multiplicity {\it fluctuations}. The probabilty is not negligible that an event may `fluctuate' so as to produce {\it no} secondaries in the trigger interval. We cannot estimate this probability assuming a Poisson distribution over the multiplicity of the final hadrons. The reason is that the hadrons are dominantly produced via minijet fragmentation and the mean number of minijets is much smaller than the
multiplicity of final hadrons. However, CDF observations \cite{cdf} indicate that the probability to produce a rapidity gap of size $\Delta \eta$ due, to a fluctuation decreases as 
\be
P(\Delta\eta)\propto {\rm exp}(-a\Delta\eta),   
\label{eq:cdf}
\ee
with $a=1.3-1.4$.

It was shown in \cite{KMR} that the number of colour flux tubes produced in high-energy $pp$ interactions has a very flat energy dependence. Indeed, the density of low $p_t$ hadrons measured in the central region is practically saturated starting from the CERN-ISR energy, see Fig. 3 of Ref.~\cite{pt}. The growth of the total multiplicity $dN/dy$ is provided by the large $p_t$ hadrons, that is, by minijet production. Therefore we have no reason to expect in (\ref{eq:cdf}) a much larger power $a$ at LHC energies than that observed by CDF at the Tevatron.

Thus we estimate the probability, $P(\Delta\eta)$, that the minimum-bias triggers MBT1, MBT2, MBT3 at the LHC, will not register an event due to fluctuations. We find that in the case of ATLAS 
using trigger MBT1 (\ref{eq:MBT1}) 
we miss 7\% of the signal events. If we require the MBT3 trigger then we miss 2$\times$7\%=14\% of events, so that with such a trigger an extra 6 mb is lost\footnote{Of course, we must also correct for experimental detector efficiencies and acceptances.} from the signal of 42.4 mb. For trigger MBT2, where it is enough to have a particle in at least one detector, the probability that we miss an event due to multiplicity fluctuations is $(7\%)^2 \simeq 0.5\%$. In this case the effects of fluctuations are negligible.
 A larger rapidity interval diminishes the probability to miss 
the event due to the multiplicity fluctuation. In particular, in the interval $3.1<\eta<6.5$ the trigger (\ref{eq:MBT1}) misses only
about 1\% of the inelastic events. 

Of course, it is desirable to evaluate both the probability of fluctuations and the role of the edge effect by Monte Carlo simulations. However, the Monte Carlo model must be consistent with the particular model of soft interactions under consideration and, moreover, provide a reliable description  of all `soft' high-energy interactions. Another possibility is to use a trigger which covers a larger rapidity interval. This would greatly diminish both the probability of fluctuations and the edge effect.

Here, it is relevant to consider the so-called zero-bias measurement planned by the CMS experiment \cite{CMS}, using their central detector, together with CASTOR (on one side)\footnote{Zero-bias measurements are also planned by ATLAS.}. To be precise, without any trigger, CMS will look for secondaries in the rapidity interval $-5<|\eta|<7$  in a `random' event in a bunch crossing. The model of \cite{KMR} predicts values of the signal given in Table \ref{tab:T3}. For completeness, we also give the values for using the central detector on its own.  We see, for example, that even if we demand one or more secondaries in the extended rapidity interval $-5<\eta<7$, then we miss\footnote{Note that it is possible to cover larger values of rapidity using the Zero Degree Calorimeters (ZDC), with coverage $|\eta|>8.5$, or the
Forward Shower Counters \cite{FSC} (FSC), with coverage $9.5<|\eta|<11$. However, the inclusion of these measurements will require a reliable MC model which accounts for the detailed structure of low-mass proton dissociation, including the diffractive production (and decay)
of the different baryon resonances.} a component of 12.4 mb from the total inelastic cross section at the LHC energy $\sqrt{s}$=14 TeV. The probability of 
fluctuations, with such an extended rapidity interval, is completely negligible, $\sim 10^{-7}$. 
\begin{table} [h]
\centering
  \begin{tabular}{c|cc|cc}
    \hline \hline
    & \multicolumn{2}{c|}{$-5<\eta<7$} & \multicolumn{2}{c}{$-5<\eta<5$} \\ \hline
     $\sqrt{s}$ & 14 TeV & 10 TeV & 14 TeV & 10 TeV \\ \hline
     
$\sigma_{{\rm low}M}$ &   26.7   &   25.7  & 26.7 & 25.7    \\
$\sigma_{1{\rm high}M}$  &   7.0  &   5.6  & 10.6 & 9.3   \\ 
$\sigma_{2{\rm high}M}$  &   0.2   &   0.1 & 0.7 & 0.5   \\ \hline
$\sigma_{\rm signal}$  &   57.6   &   57.2 & 53.5 & 53.1 \\    
    \hline \hline
  \end{tabular}
\caption{\sf The cross sections (in mb), $\sigma_{\rm signal}$, predicted using the model of \cite{KMR} for the CMS zero-bias measurements using, first, the central detector with CASTOR, covering the $-5<\eta<7$ rapidity interval, and, second, using the central detector alone, which covers the interval $-5<\eta<5$.  Again the first three lines show, similar to Tables \ref{tab:T1} and \ref{tab:T2}, the cross section components missed by the zero-bias measurement.}
  \label{tab:T3}
\end{table}

To conclude, we  emphasize that even minimum-bias triggers are only sensitive to about half of the total cross section. Therefore, it is important to present the results of the original measurements, together with the detail description of the conditions required by the trigger.
It is much more informative to compare with one or another model the actual cross sections\footnote{The minimum-bias cross section, $\sigma_{\rm MB}$, is $\sigma_{\rm signal}$ of (\ref{eq:signal}) corrected for fluctuation and edge effects.}, $\sigma_{\rm MB}$, measured using the minimum-bias triggers, rather than the 
total inelastic cross section obtained by a model-dependent extrapolation from  $\sigma_{\rm MB}$
to $\sigma_{\rm inel}$. Then one can check the model just at the  $\sigma_{\rm MB}$ level, including  the dependence of $\sigma_{\rm MB}$ on the size and the position of the rapidity
intervals covered by the detectors, e.g. to compare the results obtained with the triggers MBT1, MBT2 and MBT3.
In particular, the Tel-Aviv model \cite{GLMM}, which has a very small probability of high-mass dissociation, would predict, at $\sqrt{s}$=14 TeV, the same cross sections $\sigma_{\rm MB}\simeq 55$ mb in all three cases. Despite the fact that the total and elastic cross sections ($\sigma_{\rm tot}=92.1$ mb and $\sigma_{\rm el}=20.9$ mb) of this model are very close to those of the Durham model \cite{KMR},  we expect a noticeable difference in the values of $\sigma_{\rm MB}$.

Only a model which is found to give a satisfactory prediction for the measured  $\sigma_{\rm MB}$ can be used to extrapolate to obtain the value of $\sigma_{\rm inel}$.

\section*{Acknowledgements}

We are very grateful to Mike Albrow, Craig Buttar, Albert De Roeck, Risto Orava and Andy Pilkington for stimulating our interest in this problem.  MGR would like to thank the IPPP at the University of Durham for hospitality. This work was supported by the grant RFBR
07-02-00023, by the Federal Program of the Russian State RSGSS-3628.2008.2 and by the Russia-Israel grant 06-02-72041-MNTI.


\thebibliography{}
\bibitem{MBTS} ATLAS Collaboration: G.~Aad {\it et al.}, arXiv:0901.0512 [hep-ex].

\bibitem{TOTEM} TOTEM Collaboration: V.~Berardi {\it et al.},
CERN-LHCC-2004-002, Jan 2004; CERN-LHCC-2004-020, Jun 2004.

\bibitem{soft} M.G. Ryskin, A.D. Martin and V.A. Khoze, Eur. Phys. J. {\bf C54}, 199 (2008).

\bibitem{KMR} M.G. Ryskin, A.D. Martin and V.A. Khoze, Eur. Phys. J. {\bf C60}, 249 (2009).

\bibitem{cdf} CDF Collaboration: T. Affolder et al., Phys. Rev. Lett. {\bf 87}, 141802 (2001).

\bibitem{pt} CDF Collaboration: F. Abe et al., Phys. Rev. Lett. {\bf 61}, 1819 (1988)

\bibitem{CMS} CMS Collaboration: G.L.~Bayatian {\it et al.},  CERN-LHCC-2006-001, CMS-TDR-008-1, 2006;
  J.\ Phys.\ G {\bf 34} (2007) 995.

\bibitem{FSC} USCMS Collaboration: M.~Albrow {\it et al.},
  arXiv:0811.0120 [hep-ex].

\bibitem{GLMM} A. Gotsman, E. Levin, U. Maor and J. Miller, Eur. Phys. J. {\bf C57}, 689 (2008).
\end{document}